\documentclass[%
 aip,
 amsmath,amssymb,
 reprint,superscriptaddress,%
]{revtex4-1}

\usepackage[
	pdffitwindow=true,
	colorlinks=true,
	frenchlinks=false,
    linkcolor=blue,
    anchorcolor=blue,
    citecolor=blue,
    filecolor=blue,
    urlcolor=blue,
    bookmarks=true,
    bookmarksopen=true,
	bookmarksnumbered=true, 
    bookmarksopenlevel=1,
    plainpages=false,
	pdfpagelayout=SinglePage,
    pdfpagelabels=true,
	breaklinks
]{hyperref}
\usepackage{graphicx}% Include figure files
\usepackage{dcolumn}% Align table columns on decimal point
\usepackage{bm}% bold math
\usepackage{textcomp}
\usepackage{color}
\usepackage{siunitx}
\usepackage{soul}

\usepackage[utf8]{inputenc}
\usepackage[T1]{fontenc}
\usepackage{mathptmx}
\usepackage{soul}

\begin{document}

%\preprint{AAPM/123-QED}

\title{Initial experimental results on a superconducting-qubit reset based on photon-assisted quasiparticle tunneling}

\author{V.~A.~Sevriuk}
\affiliation{IQM, Keilaranta 19, FI-02150 Espoo, Finland}
\email{vasilii@meetiqm.com}%Lines break automatically or can be forced with \\
\author{W.~Liu}
\affiliation{IQM, Keilaranta 19, FI-02150 Espoo, Finland}
\author{J.~R\"onkk\"o}
\affiliation{IQM, Keilaranta 19, FI-02150 Espoo, Finland}
\author{H.~Hsu}
\affiliation{IQM, Nymphenburgerstr. 86, 80636 Munich, Germany}
\author{F.~Marxer}
\affiliation{IQM, Keilaranta 19, FI-02150 Espoo, Finland}
\author{T.~F.~M\"orstedt}
\affiliation{QCD Labs, QTF Centre of Excellence and InstituteQ, Department of Applied Physics, Aalto University, PO Box 13500, FI-00076 Aalto, Finland}
\author{M.~Partanen}
\affiliation{IQM, Keilaranta 19, FI-02150 Espoo, Finland}
\author{J.~R\"abin\"a}
\affiliation{IQM, Keilaranta 19, FI-02150 Espoo, Finland}
\author{M.~Venkatesh}
\affiliation{IQM, Keilaranta 19, FI-02150 Espoo, Finland}
\author{J.~Hotari}
\affiliation{IQM, Keilaranta 19, FI-02150 Espoo, Finland}
\author{L.~Gr\"onberg}
\affiliation{VTT Technical Research Centre of Finland Ltd, P.O. Box 1000, FI-02044, VTT, Finland}
\author{J.~Heinsoo}
\affiliation{IQM, Keilaranta 19, FI-02150 Espoo, Finland}
\author{T.~Li}
\affiliation{IQM, Keilaranta 19, FI-02150 Espoo, Finland}
\author{J.~Tuorila}
\affiliation{IQM, Keilaranta 19, FI-02150 Espoo, Finland}
\author{K.W.~Chan}
\affiliation{IQM, Keilaranta 19, FI-02150 Espoo, Finland}
\author{J.~Hassel}
\affiliation{IQM, Keilaranta 19, FI-02150 Espoo, Finland}
\author{K.~Y.~Tan}
\affiliation{IQM, Keilaranta 19, FI-02150 Espoo, Finland}
\author{M.~M\"ott\"onen}
\affiliation{IQM, Keilaranta 19, FI-02150 Espoo, Finland}
\affiliation{QCD Labs, QTF Centre of Excellence and InstituteQ, Department of Applied Physics, Aalto University, PO Box 13500, FI-00076 Aalto, Finland}
\affiliation{VTT Technical Research Centre of Finland Ltd, P.O. Box 1000, FI-02044, VTT, Finland}

\date{\today}% It is always \today, today,

\begin{abstract}
We present here our recent results on qubit reset scheme based on a quantum-circuit refrigerator (QCR).
In particular, we use the photon-assisted quasiparticle tunneling through a superconductor--insulator--normal-metal--insulator--superconductor junction to controllably decrease the energy relaxation time of the qubit during the QCR operation. 
In our experiment, we use a transmon qubit with dispersive readout. The QCR is capacitively coupled to the qubit through its normal-metal island. We employ rapid, square-shaped QCR control voltage pulses with durations in the range of 2--350~ns and a variety of amplitudes to optimize the reset time and fidelity. 
Consequently, we reach a qubit ground-state probability of roughly $97\%$ with 80-ns pulses starting from the first excited state. The qubit state probability is extracted from averaged readout signal, where the calibration is based of the Rabi oscillations, thus not distinguishing the residual thermal population of the qubit.
\end{abstract}

\maketitle
As formulated in the DiVincenzo criteria~\cite{Divincenzo00}, a functional quantum processor calls for the ability to initialize the qubits into a desired pure state. Seemingly, this criterion is naturally satisfied in the current state-of-the-art quantum processors due to the natural decay of the excited states of the quantum system toward its ground state. However, to increase the speed and thus the performance of the quantum processors, we need to make the initialization fast and accurate.  Depending on the application, either the whole qubit register or just parts of it has to be reset during the algorithm. For example, the whole register is initialized in variational hybrid quantum-classical algorithms~\cite{McClean2016} and only selected qubits in error correcting codes~\cite{Fowler2012} which in fact, pose stringent requirements on the reset speed.

This paper focuses on the reset of superconducting circuits by photon-assisted quasiparticle tunneling through a superconductor--insulator--normal-metal--insulator--superconductor (SINIS) junction. We report on such a quantum-circuit refrigerator (QCR)~\cite{Tan2017,morstedt2021}, a versatile tool, which has already been used for on-demand reset of a superconducting resonator~\cite{sevriuk_2019}, among its other applications~\cite{Partanen18, hyyppa2019, Silveri2019, viitanen2021,morstedt2021}. The fact that superconducting circuits are also extensively used in hybrid quantum systems~\cite{xiang2013} further widens the possible applications of the QCR.

From the fundamental point of view, the QCR acts as a widely controllable environment for quantum systems, and hence provides the possibility to study different aspects of the physics of open quantum systems~\cite{Leggett1987, Rastelli2018, pttheory2018, Clerk10, sweeney2019, Heiss2012}. For example, QCR may turn useful for dissipation-driven quantum information processing, simulation, and protection~\cite{Mirrahimi_2014,kraus2008,Verstraete2009}.

\begin{figure}
  \centering
  \includegraphics[width=1\linewidth]{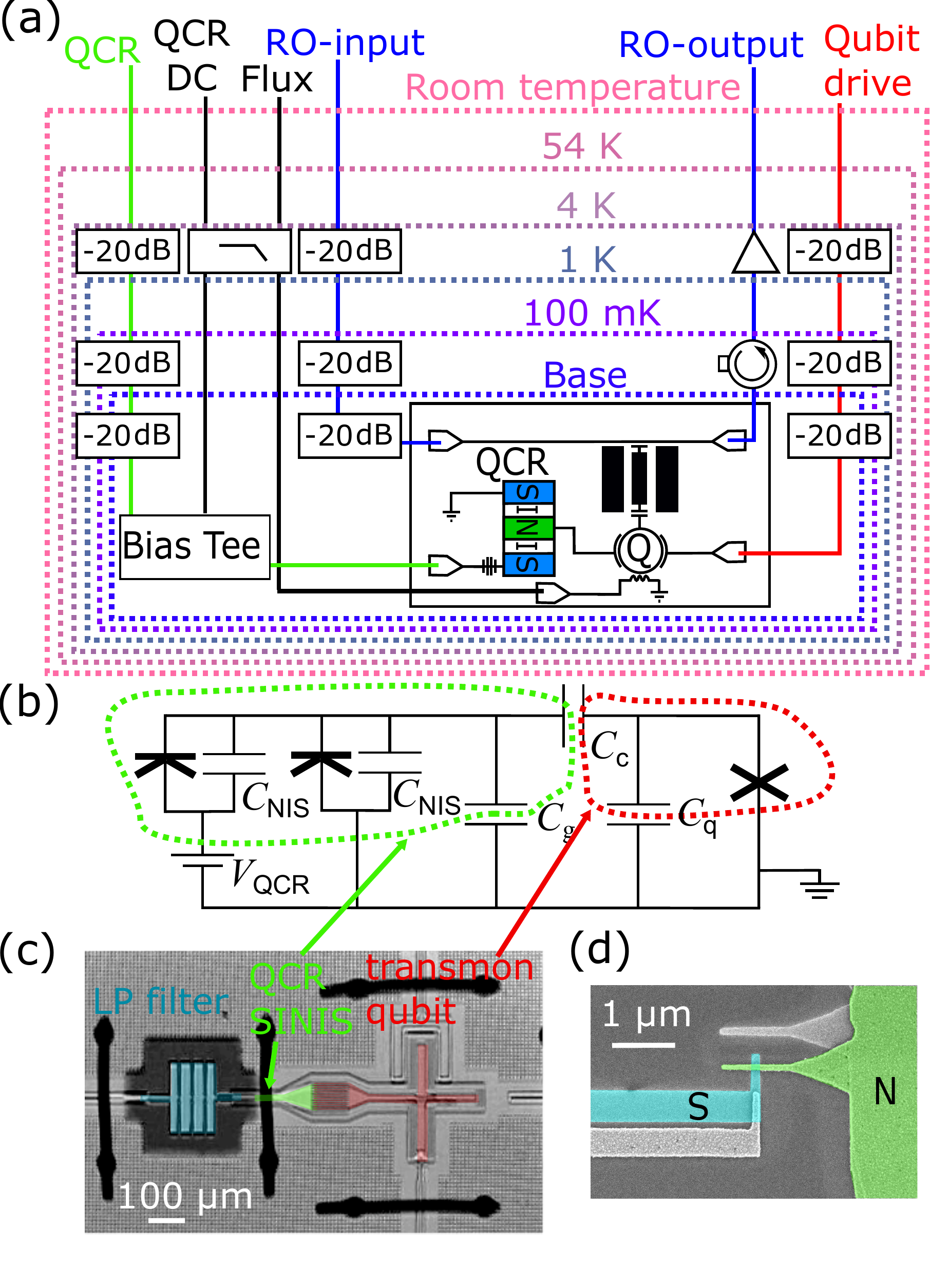}
  \caption{\label{fig:scheme}(a) Wiring scheme of the sample inside the cryostat. The drive line, readout input line, and QCR rapid pulse line are anchored at three different temperatures and have 60 dB of total attenuation each. The readout output line is equipped with a low-temperature high-electron-mobility-transistor (HEMT) amplifier. The QCR direct-current (DC) line and the flux line of the frequency-tunable qubit are twisted pairs with resistive low-pass filters. The QCR pulse and DC lines are connected to the bias tee, and the combined signal is connected to the QCR through the on-chip 0.3-GHz low-pass Pd filter, see panel (c).
  (b)~Simplified circuit diagram of the sample, excluding the readout resonator. The qubit is displayed as a Josephson junction (black symmetric cross) and the parallel capacitor $C_\mathrm{q}$.
  The QCR is depicted as two normal-metal--insulator--superconductor (NIS) junctions (black half crosses) parallel to the island-to-ground capacitor $C_\mathrm{g}$.
  The QCR control voltage is denoted by $V_\mathrm{QCR}$. The parts of the schematic corresponding to the QCR and the qubit are each encircled with colors corresponding to the parts in panel (c). (c) Colored scanning-electron-microscope image of the sample showing the qubit (red color), QCR (green color) and low-pass (LP) Pd filter (blue color). The black stripes are bonding wires connecting different parts of the ground plane, but not the other conductors. (d)~Colored scanning-electron-microscope image of the one of the QCR NIS junctions, where blue color denotes the superconductor and green stands for the normal-metal island.}
\end{figure}

\begin{figure}
  \centering
  \includegraphics[width=1\linewidth]{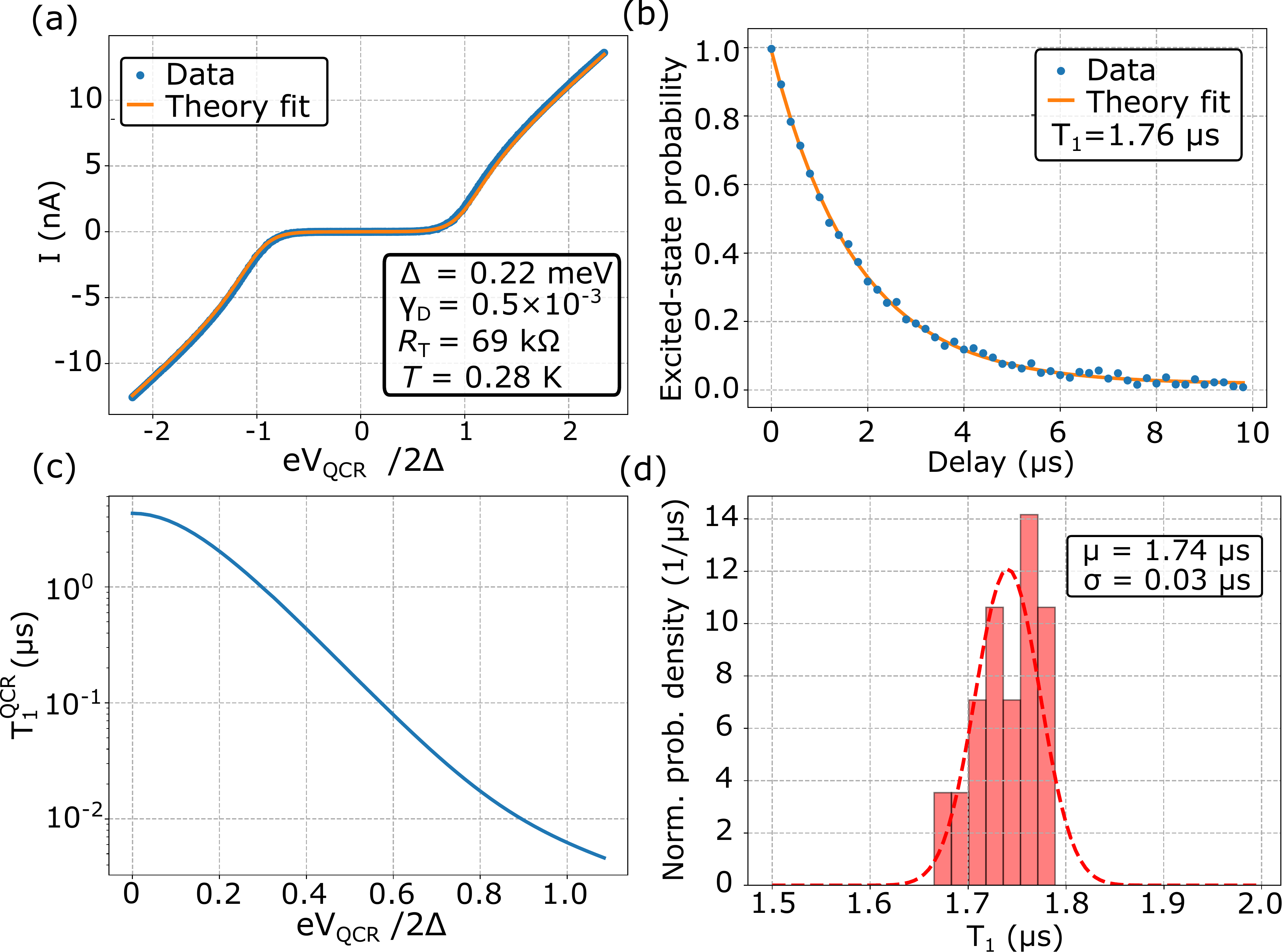}
  \caption{\label{fig:DC}(a) Example trace of the QCR DC current as a function of its DC voltage. The experimental data (blue markers) fit accurately to the conventional current-voltage characteristics of NIS junctions (orange line)~\cite{Muhonen2012}. The resulting values of the fitting parameters are indicated. (b) Example trace of the excited-state probability of the qubit as a function of time in a typical T$_1$ measurement of the qubit with no voltage applied on the QCR (QCR off-state). (c) Theoretically calculated energy decay time, T$_1^{\rm{QCR}}$, of the qubit owing to the QCR.
  The result is obtained using the model described in Refs.~\onlinecite{Silveri17prb, hsu2020} and the parameter values in Table~\ref{t1}. (d) Probability distribution of T$_1$ measurement results in the QCR off-state obtained by repeating the experiment in (b). These measurements are interleaved with the reset experiment, the pulse sequence of which is shown in Fig.~\ref{fig:result}(a).}
\end{figure}

Figure~\ref{fig:scheme}(b) shows a circuit diagram of the QCR which consists of two NIS junctions in series.
The QCR is operated by a voltage $V_\mathrm{QCR}$ applied across its superconducting electrodes, and it is capacitively coupled to the targeted circuit through its normal-metal island. The operation of the QCR is mainly determined by three processes: elastic quasiparticle tunneling through the NIS junctions and inelastic quasiparticle tunneling with photon absorption or emission. For each applied control voltage $V_\mathrm{QCR}$, there are certain rates for each of these processes, which eventually determines the speed and accuracy of the reset. During the standard operation of the QCR, the elastic tunneling is the dominant process, which defines the QCR current--voltage characteristic, see Fig.~\ref{fig:DC}(a). To tunnel into the superconducting electrode from the normal-metal island, a quasiparticle needs an additional energy equal to or higher than $\Delta$, where $2\Delta$ is the energy gap in the superconductor density of states. For the elastic tunneling, this energy is fully provided by the voltage applied across the junctions added with thermal fluctuations. Thus there is an exponential rise of the elastic tunneling rate when the QCR control voltage $V_\textrm{QCR}$ approaches $2\Delta/e$ across the SINIS structure having two junctions in series. For the inelastic tunneling the energy can be also obtained from or given to the qubit. Thus the energy constrain for the photon absorption process is relaxed by the qubit excitation energy $2\pi\hbar f_0$ with respect to the elastic tunneling, and the photon emission process demands the same amount of energy in addition to the elastic tunneling process. This leads to an optimal regime of the QCR operation at the voltage level close to the edge of the subgap regime, where the ratio between photon absorption and emission is maximized~\cite{Silveri17prb}.
Our model is based on the well-developed theory of the QCR which can be found in the previous publications~\cite{Silveri17prb, hsu2020, Hsu2021, vadimov2021, morstedt2021}.  

Previous works related to superconducting-qubit reset can be divided into a few groups. The first group utilizes fast sweeps of the qubit frequency~\cite{resetReed2010, McEwen2021}, which however may lead to undesired excitations and 
frequency crowding in multiqubit devices. Another large group is microwave control of the qubit state~\cite{resetGeerlings2013,pulseresetEgger2018,resetMagnard2018, Zhou2021,Valenzuela2006}. Part of the research in this direction is based on the conditional reset of the qubit, which requires readout of the qubit state. This naturally limits such reset protocols to the readout fidelity, readout time, and the relatively long feedback time. Nevertheless, recently prominent success was reached by unconditional reset protocol~\cite{resetMagnard2018} with a reset time of around 300~ns and residual qubit population less than 0.3\%. Similar results were also reached by modulating the flux through a transmon qubit~\cite{Zhou2021}.

In Ref.~\onlinecite{resetMagnard2018}, the reset time was limited by the photon decay rate through the resonator that is coupled to the qubit. 
Naively increasing this decay rate also leads to a decrease in the qubit lifetime. A trade-off between the reset time and qubit coherence is also relevant for our device, arising from the fact that the QCR introduces additional qubit decay channels even in the off-state, mainly because of subgap junction conductance and ohmic losses in the normal metal. Fortunately, it has been previously shown that for the QCR, the decay rates of the on- and off-state are roughly four orders of magnitude apart~\cite{sevriuk_2019}. As an additional preservation measure, we may consider a circuit where the QCR is coupled to the qubit through another supplemental qubit or resonator~\cite{hsu2020, vadimov2021}.
Such a scheme also provides an opportunity to combine the QCR-tunable decay rate of a resonator with the unconditional fast reset of the qubit by the microwave drive~\cite{resetMagnard2018, Zhou2021}. Recent theoretical work of this scenario seems promising~\cite{yoshioka2021fast}, but to date, no experimental implementation has been reported.

Here, we focus on the purely QCR-driven reset, where the QCR is directly coupled to the targeted qubit. Thus, this study falls into the area of qubit reset by rapid control of its engineered environment~\cite{Jones2013,basilewitsch2019,Tuorila2017,Wong_2019}.

The samples, illustrated in Fig.~\ref{fig:scheme}(c), are fabricated at the OtaNano Micronova cleanrooms. First, a high purity 200-nm-thick Nb layer is deposited on a high-resistivity ($\rho$  > 10 k$\Omega$ cm) non-oxidized $n$-type undoped (100) 6-inch silicon wafer by sputtering. Then, coplanar waveguides and capacitor structures are formed by photolithography with subsequent reactive ion etching (RIE). After etching, the photoresist residuals are cleaned in an ultrasonic bath with acetone and isopropanol (IPA). Next, a 45-nm-thick film of dielectric $\mathrm{Al_{2}O_{3}}$ is grown by atomic-layer deposition (ALD).  
In the subsequent photolithography step, this dielectric layer is wet-etched away from everywhere except from the location of parallel-plate ground capacitors of a low-pass filter. 
We also use electron beam lithography (EBL) with subsequent e-beam evaporations and a lift-off processes to form the low-pass filter, the QCR, and the Josephson junctions of the qubit. For the low-pass filter we deposit a 30-nm Pd layer.
Two-angle evaporation is used for both QCR and SIS junctions.
The QCR is formed by 20 nm of Al and 40 nm of Cu. The qubit junctions are formed by two 20-nm Al layers. In both structures, the metal layers are separated by a dielectric layer of $\mathrm{Al_{x}O_{y}}$ formed by in-situ oxidation. 
Before the evaporation of these structures natural oxides are removed from the surface by Ar ion milling. More information about the QCR fabrication can be found in Ref.~\onlinecite{Tan2017}. The low-pass filter mentioned above is designed as a lumped-element resistor-capacitor (RC) circuit, where the capacitors are formed by the wide fins connected by the narrow bridges, which operate as the resistors. Based on our initial estimates, the electron temperature of the filter is essentially unchanged for a single QCR voltage pulse and elevated by up to a few tens of millikelvins if QCR was turned on and off every microsecond.
The resulting sample parameters are given in Table~\ref{t1}.

The experimental setup including the main wiring scheme is shown in Fig.~\ref{fig:scheme}(a).
We conduct our experiment in a Bluefors dilution refrigerator with a base temperature below 10~mK. We employ a conventional flux-tunable transmon qubit with a capacitively coupled drive line and a resonator for dispersive readout~\cite{Blais04}. The flux-tunability is not used in the current experiment and we carry out the measurements at the flux sweet spot, where the qubit frequency reaches its maximum. 

\renewcommand{\arraystretch}{1.5}
\begin{table}
  
  \caption{\label{t1}Key parameters of the measured sample together with their descriptions.}
  \begin{ruledtabular}

 \begin{tabular}{c c c}
  
    $R_\mathrm{T}^\mathrm{NIS}$ & $34.5$ $\mathrm{k\Omega}$ & NIS tunneling resistance ($R_\mathrm{T}^\mathrm{NIS}=R_\mathrm{T}/2$)  \\ 
       \hline 
    $T_\mathrm{N}$ & 0.28 K & electron temperature of the normal-metal island \\ 
       \hline        
    $\gamma_\mathrm{D}$ & $5\times10^{-4}$ & Dynes parameter \\ 
       \hline 
    $Z_\mathrm{r}$ & 179  $\mathrm{\Omega}$ & qubit characteristic impedance \\ 
       \hline 
    $C_\mathrm{q}$ & 97 fF & qubit capacitance  \\ 
       \hline
    $C_\mathrm{c}$ & $15 $ fF & capacitance between QCR and qubit  \\ 
       \hline 
    $C_\mathrm{NIS}$ & $3.5 $ fF & NIS junction capacitance \\ 
       \hline
    $C_\mathrm{g}$ & $24.4 $ fF & QCR normal-metal-island capacitance \\ 
       \hline 
    $f_\mathrm{0}$ & 9.18 GHz & qubit frequency \\ 
       \hline 
    $\mathrm{\Delta}$ & 220 $\si{\micro\eV}$ & energy gap parameter of the Al leads \\ 
      
  \end{tabular}
  \end{ruledtabular}

\end{table}

In our experiments, we begin with the characterization of the qubit and the QCR separately, which includes conventional qubit characterization~\cite{Krantz2019} and DC measurements of the QCR junctions illustrated in Fig.~\ref{fig:DC}. From these experiments, we estimate the NIS tunneling resistance, the normal-metal electron temperature, the Dynes parameter, the superconductor energy gap parameter, and the qubit capacitance, frequency, and characteristic impedance given in Table~\ref{t1}. We make these estimates based on the assumption of symmetric junctions. The junction capacitance and QCR--qubit coupling capacitance are estimated based on electromagnetic modeling. Importantly the normal-metal electron temperature is expected to be dependent on the QCR control voltage. However the fitting of the QCR DC current-voltage curve yields a single characteristic value. The dedicated study of the temperature dependence on the applied voltage is left for the future work.

With the obtained parameter values, we use our theoretical model~\cite{Silveri17prb, hsu2020} to show in Fig.~\ref{fig:DC}(c) the energy relaxation time, T$_1^{\rm{QCR}}=1/[\Gamma_{10}(V_{\rm{QCR}})+\Gamma_{01}(V_{\rm{QCR}})]$, of the qubit at different fixed QCR control voltages. Expressions for the QCR-induced relaxation $\Gamma_{10}(V_{\rm{QCR}})$ and excitation $\Gamma_{01}(V_{\rm{QCR}})$ rates are given in Ref.~\onlinecite{hsu2020}. 
From our measurements of similar samples without a QCR, we have obtained energy decay times greatly exceeding the
T$_1^{\rm{QCR}}$ of 4.31 $\si{\micro\s}$ predicted from the QCR in the off-state.
However, the actual measured qubit T$_1$ in the QCR off-state is 1.74~$\pm{}$~0.033~$\si{\micro\s}$, which is substantially lower than our prediction, see Fig.~\ref{fig:DC}(b,d). Subsequently, we carried out classical electromagnetic simulations which showed that this discrepancy can be explained by ohmic losses arising from the normal-metal island in close proximity to the qubit. The simulations are showing that in our current design, the qubit electromagnetic field is coupling to the normal-metal island through the QCR--qubit capacitor as well as being mediated by the ground plane metal and a direct parasitic capacitance. Fortunately, these losses can be minimized in a future redesign of the sample. Another possible source of qubit decoherence arises from the Purcell decay of the qubit to the QCR control line. We estimate that in our experiment, it limits the qubit T$_1$ time to roughly 12 $\si{\micro\s}$. This value is significantly higher than that in our experiments and than the theoretical limit for the current QCR parameters.  The on-chip Pd filter between the QCR and the 50-$\Omega$ control line makes this decoherence source insignificant in our current discussion.  

Figure~\ref{fig:DC}(c) suggests that it is possible to use the QCR to tune the qubit T$_1$ by a factor of roughly 1000, from microseconds to nanoseconds, which can be beneficial in the practical applications discussed above. However, we need to first demonstrate, how such T$_1$ tuning works in the case of rapid switching between these two regimes, and to find optimized pulse parameters for the QCR control voltage.

For the time domain experiments we used averaged readout. We first calibrate our qubit drive pulse and readout with a typical Rabi experiment~\cite{Krantz2019}. We establish the parameters of the qubit drive for a $\pi$ pulse to fully excite the qubit and define the position of the excited and the ground state in the in-phase--quadrature-phase (IQ) plane of the down-converted readout pulse~\cite{naghiloo2019}. We use 2-$\si{\micro\s}$ readout pulses and 120-ns $\pi$ pulses, and average over 32768 single trace measurements to extract the qubit excited-state probability with fixed QCR control parameters.
All our following results are presented in the terms of the qubit excited-state probability. It means also that the qubit decay during the $\pi$ pulse and the readout cancels in the process of the readout calibration which is based on the Rabi experiment.

\begin{figure}
  \centering
  \includegraphics[width=0.9\linewidth]{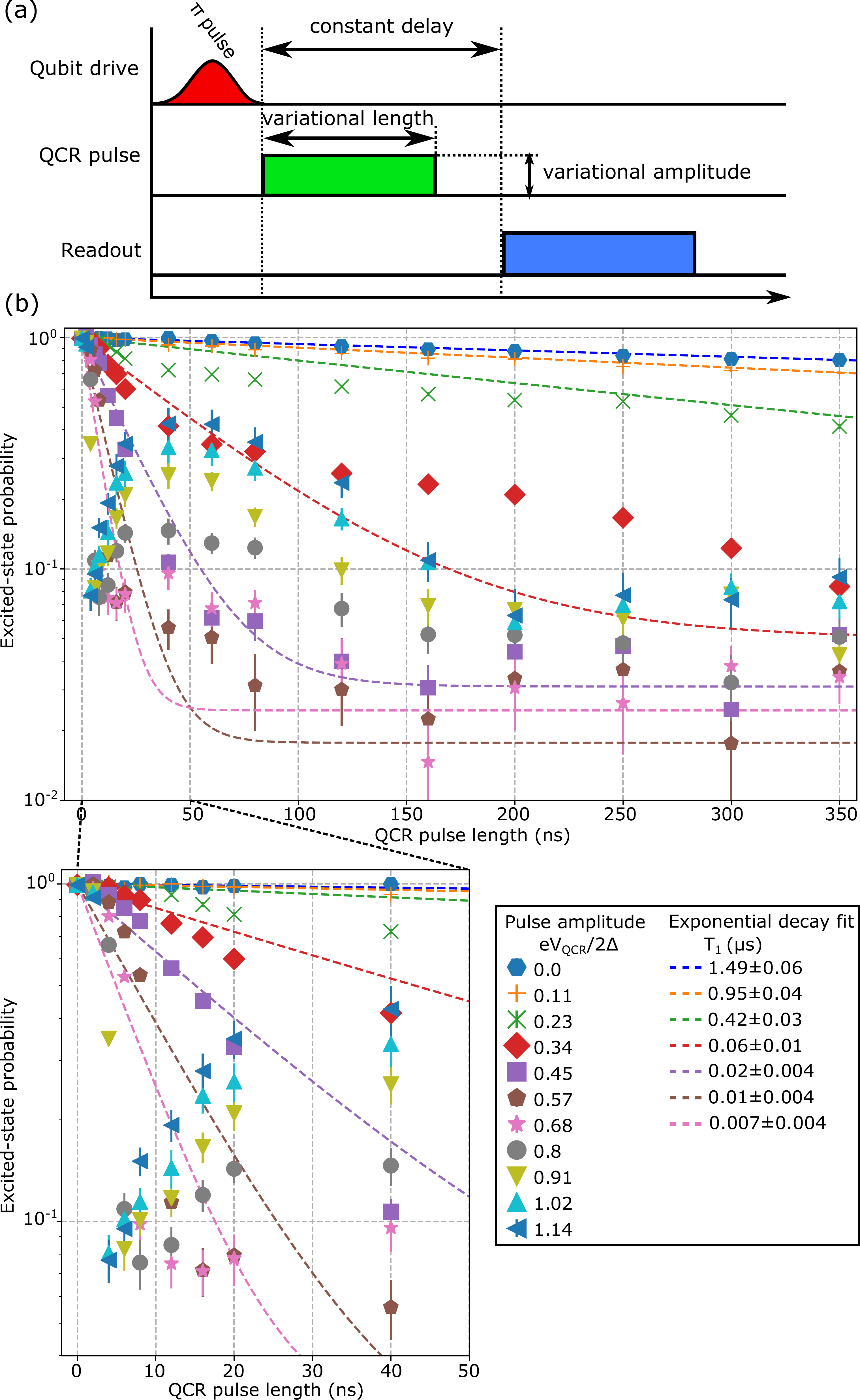}
  \caption{\label{fig:result}(a) The measurement protocol where at first, the qubit is driven to the maximum probability of its excited state by a $\pi$ pulse, shown in the top row. Then, a QCR pulse is sent during the delay between the drive and readout pulses, shown in the middle row. The QCR pulse amplitude and length is varied such that the QCR pulse is never overlapping with the drive or readout pulse applied at the end (bottom row) of a single realization of the protocol.
  (b) Excited-state probability of the qubit as a function of the length of the QCR voltage pulse of different indicated amplitudes. Zero-length data is obtained by separate interleaved measurements, where the readout is applied immediately after the drive pulse. For the other points, we employ the pulse sequence of  panel~(a). The excited-state probability is calibrated by a Rabi experiment before the measurements. For the lowest six amplitudes, exponential fits are shown by dashed lines. For the rest of the amplitudes, such fitting is not reasonable owing to the additional raise of the excited-state probability after the initial fall. The inset shows a magnification to the pulse length range 0--50 ns.}
\end{figure}

The pulsing scheme of our time domain experiment is shown in Fig.~\ref{fig:result}(a). First, we excite the qubit with a $\pi$ pulse, then we send a rectangular voltage pulse to the QCR, which is aligned to the end of the excitation pulse, and finally apply a qubit readout pulse.
We vary the length and the amplitude of the QCR pulse to study the  decay of the qubit excitation. Importantly, the delay between the drive and readout pulses is fixed at 400 ns and long enough to fit the longest QCR pulse in between without temporal overlap. For each QCR pulse amplitude, we extract the contribution of the different pulse lengths to the qubit excited-state probability by first subtracting the effect of the natural decay of the qubit during the time between the $\pi$ pulse and readout when the QCR pulse is off. Interleaved with these experiments, we also measure the qubit state without any delay between the $\pi$ pulse and readout, as a part of our T$_1$ measurements, see Figs.~\ref{fig:DC}(b) and~\ref{fig:DC}(d).

Figure~\ref{fig:result}(b) shows the measured excited-state probability of the qubit as a function of the QCR pulse length for different QCR pulse amplitudes.
Except for the couple of lowest amplitudes, the decay is not well described by a single exponential as expected from an ideal model. This can be attributed to a few different effects: Firstly, the bias tee at the base temperature works as a high pass filter for the QCR square pulses. We estimate the time constant of the bias tee around 500 ns.
Secondly, the charging of the QCR normal-metal island takes place through the nonlinear resistance of the junctions, which can lead to the complicated and individual effective pulse distortions at each junction. See the Supplementary material for the further details. At high pulse amplitudes we observe an increase in the excited-state probability after an initial fast drop, which can be explained by the fact that the QCR is starting to excite higher levels of the qubit, which leads to the signal drifting out of its usual distribution in the IQ plane~\cite{resetMagnard2018}. Due to this our data analysis, which neglects the higher qubit states, is leading also to increased uncertainty.

Since our data show a more complex behavior than a simple exponential decay, it is perhaps  more informative to study the shortest time to achieve the minimal excited-state probability than the decay rates. To this end, our unconditional reset can reduce the qubit excited-state probability from nearly 100$\%$ to $(3~\pm{}~1)\%$ in 80 ns with the QCR control voltage pulse amplitude of 0.57$\times 2\Delta$. The numbers here are based only on the Rabi experiment and thus does not include the residual thermal population of the qubit. With the current QCR parameters our theoretical model predicts that the residual thermal population of the qubit at the QCR off-state is roughly 5\%{}.

A simple exponential fit yields T$_1$ = 10~$\pm{}$~4 ns at a pulse amplitude of 0.57$\times 2\Delta$. This time is 1/170 times the qubit T$_1$ time in the QCR off-state. In the future, this on/off ratio can be increased by a new design of the coupling between the QCR and the qubit, in which the transmission line between the QCR and the qubit will increase distance and hence decreases the ohmic losses from the normal-metal island~\cite{vadimov2021}. This will increase T$_1$ in the QCR off-state likely close to its theoretical value of 4.31~$\si{\micro\s}$ given the parameters of Table~\ref{t1}. Furthermore, an additional filter may reduce the excitation of the high-energy qubit states, rendering high QCR pulse amplitudes more effective. From our model, we predict that in such a case, T$_1$ drops to approximately 7~ns when the QCR is on. Hence, these two future modifications potentially lead to a QCR on/off  ratio of a thousand. Theoretically~\cite{hsu2020}, the static on/off ratio is approximately given by $\sqrt{\Delta/(2\pi\hbar f_0\gamma_\mathrm{D}^2)}\approx 5000$, at low electron temperatures and $f_0\ll2\pi\Delta/\hbar$.

According to our theoretical model, by properly choosing the tunneling resistance $R_\mathrm{T}$ of the QCR and the coupling strength, we can scale the overall dissipative effect of the QCR on the qubit, and hence move to a regime where the QCR will not be a limiting factor for the qubit T$_1$ in the off-state. Thus if the off-state T$_1$ is 50~$\si{\micro\s}$, the QCR may allow to switch it to roughly 50~ns or below for a short period of time. Such a regime can be of practical value in quantum information processing, and thus motivates future research on the QCR.

Further improvements of these numbers seem to some extent feasible, for example, with the reduction of the Dynes parameter and the effective temperature of the QCR normal-metal island, which calls for a combination of specifically designed heat sinks and advanced on-chip filters~\cite{Muhonen2012}. Recent achievements in the tunnel junction thermometry show temperatures of the NIS junction down to the single-millikelvin regime~\cite{Feshchenko2015} with the range of 10$^{-7}$ for the Dynes parameter~\cite{Saira2012}. Another approach is to combine the QCR with other techniques such as flux-tunable resonators~\cite{Partanen18,yoshioka2021fast} or driving schemes~\cite{viitanen2021}.

For potential future industrial applications, the aging of the QCRs needs to be considered. Although Cu is stable at low temperatures, it can oxidize and degrade when exposed to atmosphere or thermal cycling \cite{Choudhary2018}. The room temperature QCR junction resistance has shown to increase significantly within weeks at atmosphere, calling for improved storage techniques, e.g., nitrogen freezers, or alternative normal-metal materials~\cite{Muhonen2009,Santos2016} such as AuPd or AlMn.

On one hand, additional integration efforts are needed to utilize QCR in practical quantum-information-processing devices due to the need of extra control lines and noise mitigation. On the other hand, the form factor of the QCR is small and it is exponentially tuned by a simple voltage pulse. Such simplicity is an advantage of the QCR in combination with its ability of unconditional qubit reset. In view of future scaling of the quantum processors in the number of qubits, we emphasise that a single QCR can be used to reset multiple quantum circuits~\cite{hsu2020}. Furthermore, a single control line can be used to control multiple QCRs. 
Importantly, QCR can be perceived as a general concept for controlled engineered quantum environments, which can give rise to new quantum algorithms and protocols and help also to advance quantum simulations and the studies of fundamental physics~\cite{Silveri2019,Murch2012,Zoller1998}.

In summary, we report here the first experimental results on the utilization of the QCR for qubit reset. In the current experiment, 
the SINIS junction of the QCR is directly coupled to the qubit through a capacitor. We use simple square voltage pulses to control the relaxation time of the qubit, achieving qubit reset down to 3\% excited-state probability in 80 ns. Further work is needed to improve these numbers and reduce the dissipative effect of the QCR in its off-state on the qubit. This can be implemented by adjusting the QCR parameters and by an advanced design of the sample with additional filtering between the qubit and the QCR.

\subsection*{Supplementary Material}

See supplementary material for the cursory study of the QCR control pulse distortion due to the non-linear behavior of the NIS junctions.

\section*{ACKNOWLEDGMENTS}

We acknowledge the provision
of facilities and technical support by Aalto University at
OtaNano - Micronova Nanofabrication Centre. We have received funding from the European Research Council under grants 681311 (QUESS), 957440 (SCAR), and 101053801 (ConceptQ), European Commission through H2020 program projects QMiCS (grant agreement 820505, Quantum Flagship), Business Finland (Quantum Technologies Industrial grant No. 41419/31/2020), the Academy of Finland through its Centers of Excellence Program (project Nos. 312300, and 336810), and the Research Impact Foundation.

\section*{Data availability statement}

The data that support the findings of this study are available from the corresponding author upon reasonable request.

%\bibliography{apssamp.bib}{}
%

\section*{Supplementary information} % the title of your chapter
\subsection*{QCR control pulse distortion at the NIS junctions}

We use the simplified circuit diagram of the QCR presented in the main article to study an effective control voltage pulse distortion at the QCR junctions as shown in Fig.~\ref{fig}(a). The results of this study do not quantitatively explain our experimental data, possibly owing to oversimplification of the circuit diagram. Nevertheless, we show here  some examples as a starting point for a possible in-depth research in the future, and as a qualitative argument for the expected complex decay dynamics of the qubit population under a square voltage pulse on the QCR.
We consider a square pulse applied by the voltage source depicted in Fig.~\ref{fig}(a) with five different amplitudes and numerically model the voltage drop across each NIS junction. The resulting voltage drops are shown in Fig.~\ref{fig}(b). We also calculate qubit T$_\mathrm{1}$ for each pair of the voltages (at each point in time), shown in Fig.~\ref{fig}(c), to provide an estimation on the possible effect of the pulse distortion on the qubit.

\renewcommand{\thefigure}{S\arabic{figure}}
\begin{figure}
\includegraphics[width=0.9\linewidth]{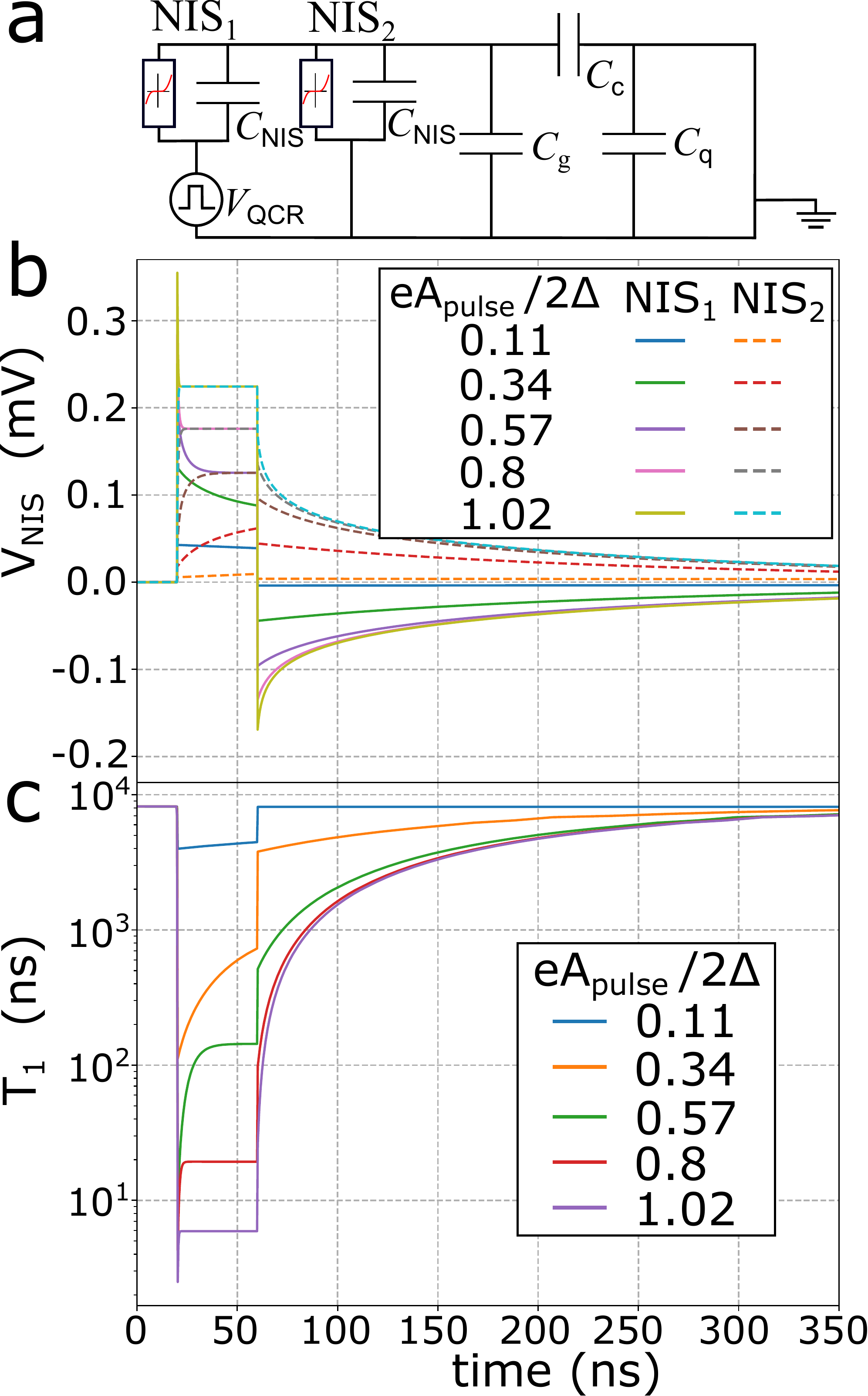}
  \caption{\label{fig}(a) Circuit diagram of the sample, which was used for the pulse distortion studies. Nonlinear resistors are marked with IV curve symbol, their parameters are deducted from the IV curve presented in the main article with the assumption that the junctions are identical. The relevant simulation parameters can be found in Table~I of the main article. (b) Voltage drops across the first and second NIS junctions [NIS$_\mathrm{1}$ and NIS$_\mathrm{2}$ as indicated in (a)] as functions of time. The voltage drops are result in from the square-shaped voltage pulse which is sent by the source source ($V_\textrm{QCR}$) depicted in panel~(a) with five different amplitudes scaled to superconductor energy gap (2$\mathrm{\Delta{}}$). The pulse length is 40~ns, and the start of the pulse is at the time instant $t=20$~ns. (c) Instantaneous T$_\mathrm{1}$ of the qubit a function of time obtained by adding the decay rates from both NIS junctions according to the voltage drops shown in panel~(b).}
\end{figure}

\end{document}